\documentclass[12pt]{iopart}
\usepackage{graphicx}
\usepackage{dcolumn}
\usepackage{amssymb}
\usepackage{bm}

\begin{document}

\title{The Proca-field in Loop Quantum Gravity}
\author{G Helesfai}
\address{\dag\ Institute for Theoretical Physics, E\"otv\"os University,
     P\'azm\'any P\'eter s\'et\'any 1/A, H-1117 Budapest, Hungary}

\eads{\mailto{heles@manna.elte.hu}}

\begin{abstract}
In this paper we investigate the Proca-field in the framework of Loop Quantum 
Gravity. It turns out that the methods developed there can 
be applied to the symplectically embedded Proca-field, giving a rigorous, 
consistent, non-perturbative quantization of the theory. This can be achieved 
by introducing a scalar field, which has completely different properties than 
the one used in spontaneous symmetry breaking. The analysis of the kernel of 
the Hamiltonian suggests that the mass term in the quantum theory has a 
different role than in the classical theory.
\end{abstract}

%\submitto{\CQG}

\maketitle

\section{Introduction}
Though loop quantum gravity (LQG) is one of the most successful non-perturbative, 
covariant theories for quantum general relativity, there is still the question 
whether its classical limit gives back general relativity ($\bar{h}\to 0$) or 
quantum field theory ($G\to 0$). For this reason we investigate a question that 
has been answered in quantum field theory, namely the case of the massive vector 
field. In quantum field theory mass is generated via spontaneous symmetry breaking due to the fact that the mass term 
in the Lagrangian of the massive vector field (the Proca-field) is 
non-renormalizable. On the other hand, LQG is a non-perturbative theory, 
therefor the question arises how to quantize this theory in the framework of 
LQG and how are the results related to those of the quantum field theory . In this paper we 
will show that the Proca-field can be quantized in a non-perturbative, 
diffeomorphism covariant way. We will see that just as in quantum field theory, a scalar 
field enters the formalism, but this scalar field will play a completely 
different role in LQG.\\
The paper is organized as follows. In section 2, we will perform the 3+1 
decomposition of the Proca-field in curved space-time and in section 3 we will 
quantize the theory using the tools developed in LQG. In section 4 we will 
discuss the results which will be summarized in the end of the paper. 
\section{Classical theory}
The action of the Proca-field coupled to gravity has the form
\begin{eqnarray}
S=\int d^4x\mathcal{L}\nonumber\\
\mathcal{L}=\mathcal{L}_G+\sqrt{-g}g^{ac}g^{bd}\Big[-\frac{1}{4}\underline{F}^4_{ab}\underline{F}^4_{cd}-\frac{1}{2}m^2 g_{ab}\underline{A}^4_c\underline{A}^4_d\Big],\label{orig}
\end{eqnarray}
where $\mathcal{L}_G$ is the Lagrangian density of the gravitational field, $g^{ab}$ is the metric-tensor, 
$g$ is its determinant and $\underline{A}^4_a$ is a $U(1)$ connection with curvature $\underline{F}^4_{ab}$. 
To apply the framework of loop quantum gravity to this system, first we have to 
do a 3+1 decomposition: introduce on the space-time manifold $M$ 
a smooth function $t$ whose gradient is nowhere vanishing and a vector field 
$t^a$ with affine parameter $t$ satisfying $t^a\nabla_a t=1$. This gives a 
foliation of space-time, i. e. each $t$ defines a 3-dimensional hypersurfice 
$\Sigma_t$. Let us decompose $t^a$ into its normal and tangential part
\begin{eqnarray}
t_a=Nn_a+N_a,
\end{eqnarray}
where $n_a$ is the unit normal of the hypersurfice $\Sigma_t$, $N$ is the 
lapse function, $N_a$ is the shift vector. Define the induced, positive-definite 
metric on $\Sigma_t$ via
\begin{eqnarray}
q_{ab}=g_{ab}+n_a n_b\label{3and1}
\end{eqnarray}
Now we use \eref{3and1} and express $n_a$ with $t_a$ and $N$, exploit 
the fact that $\sqrt{-g}=N\sqrt{q}$ and introduce the pull-backs of the 
quantities $\underline{A}^4_a\ ,\ \mathcal{D}^4_a$ to 
$\Sigma_t$ respectively $\underline{A}_a=q^b_a \underline{A}^4_b\ ,\ \mathcal{D}_a=q^b_a\mathcal{D}^4_b$
and define $A_0=t^a \underline{A}^4_a, A_0^i=t^a A^{i4}_a$. The canonical momenta are the following:
\begin{eqnarray}
\Pi_0=\frac{\delta S}{\delta \dot{A}_0}=0\quad,\quad\Pi^i_0=\frac{\delta S}{\delta \dot{A}_0^i}=0\\
\Pi_N=\frac{\delta S}{\delta \dot{N}}=0\quad,\quad\Pi_a=\frac{\delta S}{\delta \dot{N}_a}=0\\
E_a^i=\frac{\delta S}{\delta \dot{A}_a^i}\quad,\quad\underline{E}_a=\frac{\delta S}{\delta \underline{\dot{A}}_a}
\end{eqnarray}
,so we have primary constraints $\Pi_0,\Pi^i_0,\Pi_N,\Pi_a$. Putting everything together
(for details see \cite{newvar},\cite{3+1},\cite{reval}) we obtain for the Hamiltonian of the Proca-field
\begin{eqnarray}
H=\int_{\Sigma}\Big(N\mathcal{H}+N^a\mathcal{H}_a+A_0^i G_i+A_0\underline{G}+\nonumber\\
+\sqrt{q}\frac{m^2}{2}(-\frac{A^2_0}{N^2}+\frac{2}{N^2}A_0 N^b \underline{A}_b-\frac{1}{N^2}(N^b \underline{A}_b)^2)\Big)\label{origH}\\
\mathcal{H}=\frac{1}{\kappa\sqrt{q}}tr(2[K_a,K_b]-F_{ab})[E_a,E_b]+\frac{q_{ab}}{2\sqrt{q}}(\underline{E}^a\underline{E}^b+\underline{B}^a\underline{B}^b)+\nonumber\\
+\frac{\sqrt{q}m^2}{2}q^{ab}\underline{A}_a\underline{A}_b\\
\mathcal{H}_a=F_{ab}^j E^b_j+\epsilon_{abc}\underline{E}^b\underline{B}^c\\
\underline{G}=\mathcal{D}_a\underline{E}^a\\
G_i=\mathcal{D}_a E^a_i,
\end{eqnarray}
where $F_{ab}$ is the curvature of the SU(2) connection $A_a^i$, $K_a$ is 
the extrinsic curvature and $\underline{B}^a$ is the dual of the Maxwell connection $\underline{A}_a$. 
The phase space carries a symplectic structure where the (nontrivial) Poisson-brackets are:
\begin{eqnarray}
\{A_a^i(x),E_j^b(y)\}=\delta_a^b\delta_j^i\delta(x-y)\\
\{\underline{A}_a(x),\underline{E}^b(y)\}=\delta^b_a\delta(x-y)
\end{eqnarray}
Since $\dot{\Pi}_0=0$ etc. must hold, we get the following consistency conditions
(secondary constraints):
\begin{eqnarray}
0=\{\Pi_N,H\}=\mathcal{H}+\frac{1}{2}\sqrt{q}m^2\tilde{A}^2:=\tilde{\mathcal{H}}\label{co1}\\
0=\{\Pi_a,H\}=\mathcal{H}_a+\sqrt{q}m^2\tilde{A}\underline{A}_a:=\tilde{\mathcal{H}}_a\\
0=\{\Pi_0,H\}=\underline{G}-\sqrt{q}m^2\tilde{A}:=\tilde{\underline{G}}\label{co3}\\
0=\{\Pi_0^i,H\}=G_i,
\end{eqnarray}
where we used the notation $\tilde{A}=\frac{A_0-N^a\underline{A}_a}{N}$.
One can verify that the above constraints have a second class constraint algebra. To 
show this, first let us introduce the following linear combination of the primary 
constraints:
\begin{eqnarray}
\tilde{\Pi}_0:=\Pi_N+\tilde{A}\Pi_0\label{comb1}\\
\tilde{\Pi}_N:=\Pi_N+N^a\Pi_a+A_0\Pi_0\label{comb2}\\
\tilde{\Pi}_a:=\Pi_a+\underline{A}_a\Pi_0\label{comb3}
\end{eqnarray}
It is easy to show that the above linear combinations have weakly vanishing 
Poisson-brackets with all constraints, so these are first class constraints. 
The constraints $G_i$ (gravitational Gauss constraint) and $\tilde{\mathcal{H}}_a+\underline{A}_a\tilde{\underline{G}}+A_a^i G_i$ 
(diffeomorphism constraint) are also first class (as in the m=0 case, the first generates infinitesimal 
SU(2) gauge transformations, while the latter generates infinitesimal spacial diffeomorphisms). There 
are only two second class constraints: $\tilde{\underline{G}}$ and $\tilde{\mathcal{H}}$. Their Poisson-bracket is
\begin{eqnarray}
\hspace{-2cm}\{\tilde{\mathcal{H}}(x),\tilde{\underline{G}}(y)\}=\nonumber\\
\hspace{-2cm}=m^2\frac{\underline{E}^a(x) N_a(x)}{N(y)}\frac{\sqrt{q}(y)}{\sqrt{q}(x)}\delta(x-y)-\sqrt{q(x)}m^2\Big(\underline{A}^a(x)-\frac{\tilde{A}(x)N^a(x)}{N(x)}\Big)\mathcal{D}_a^{(y)}\delta(x-y):=\nonumber\\
\hspace{-2cm}:=M(x,y),
\end{eqnarray}
where $\mathcal{D}_a^{(y)}$ means that the derivative should be calculated in the y variable.\\
To deal with second class systems, one needs to introduce the so called Dirac-brackets 
instead of the Poisson brackets. In the case of field theories, it is done in the 
following way (see \cite{weinberg} for details): first one calculates the matrix 
$M_{ij}(x,y):=\{C_i(x),C_j(y)\}$, where $C_i(x)$ are the second class constraints 
in the theory. After that one calculates the inverse of $M_{ij}(x,y)$ in the following 
sense (since $M_{ij}(x,y)$ is a distribution):
\begin{eqnarray}
\int d^3 z M_{ik}(x,z)(M^{(-1)})_{kj}(z,y)=\delta_{ij}\delta(x-y)
\end{eqnarray}
After this the Dirac-bracket is defined as
\begin{eqnarray}
\{f,g\}_D:=\{f,g\}-\int d^3 x d^3 y\{f,C_i(x)\}(M^{(-1)})_{ij}(x,y)\{C_j(y),g\}\label{Diracbracket}
\end{eqnarray}
In our case, $M_{ij}(x,y)$ is a 2 by 2 matrix with components
\begin{eqnarray}
M_{11}(x,y)=M_{22}(x,y)=0\nonumber\\
M_{12}(x,y)=-M_{21}(y,x)=M(x,y)
\end{eqnarray}
The inverse of this matrix has the same structure:
\begin{eqnarray}
(M^{(-1)})_{11}(x,y)=(M^{(-1)})_{22}(x,y)=0\nonumber\\
(M^{(-1)})_{12}(x,y)=-(M^{(-1)})_{21}(y,x)=\tilde{M}(x,y),
\end{eqnarray}
where $\tilde{M}(x,y)$ satisfies the following differential equation
\begin{eqnarray}
\hspace{-2cm}m^2\frac{\underline{E}^a(x) N_a(x)}{N(y)}\sqrt{q}(x)\tilde{M}(x,y)-\sqrt{q(x)}m^2\Big(\underline{A}^a(x)-\frac{\tilde{A}(x)N^a(x)}{N(x)}\Big)\mathcal{D}_a^{(x)}\tilde{M}(x,y)=\delta(x-y),\nonumber\\
\end{eqnarray}
thus the Dirac-bracket has the form
\begin{eqnarray}
\hspace{-2cm}\{f,g\}_D:=\{f,g\}+\int d^3 x d^3 y\Big(\{f,\tilde{\mathcal{H}}(x)\}\tilde{M}(y,x)\{\tilde{G}(y),g\}-\{f,\tilde{G}(x)\}\tilde{M}(x,y)\{\tilde{\mathcal{H}}(y),g\}\Big)\nonumber\\
\end{eqnarray}

The above construction has two drawbacks: the first is that the above Lagrangian is not gauge invariant. 
The cause of this is the mass term, since if one replaces m=0, we will have a first 
class constraint algebra. This affects only the U(1) gauge, SU(2) symmetries are still valid (since the 
mass term contains gravitational variables in the form of the metric tensor and its determinant). 
The second problem is that one is lead to a system with second-class constraints. 
The latter problem is more troublesome because the canonical quantization 
becomes quite difficult due to the fact that it is far nontrivial to implement 
the Dirac-brackets in the quantum theory (\cite{constr}).\\
There is an elegant way of curing both problems (\cite{constr},\cite{cl},\cite{ccl}), and 
that is to introduce an auxiliary scalar-field and modify the Lagrangian to 
have the following form:
\begin{eqnarray}
\mathcal{L}_m=\mathcal{L}_G+\sqrt{-g}g^{ac}g^{bd}\Big[-\frac{1}{4}\underline{F}^4_{ab}\underline{F}^4_{cd}-\frac{1}{2}m^2 g_{ab}(\underline{A}^4_c+\partial^4_c\phi)(\underline{A}^4_d+\partial^4_d\phi)\Big]=\nonumber\\
=\mathcal{L}_G+\mathcal{L}_{YM}+\mathcal{L}_M\label{sympl}
\end{eqnarray}
Note that the above Lagrangian is gauge invariant if the transformation rule 
for the fields under gauge transformation is
\begin{eqnarray}
\delta\underline{A}^4_{\mu}=\partial^4_{\mu}\Lambda\\
\delta\phi=\Lambda
\end{eqnarray}
and the original Lagrangian is obtained via the gauge-fixing $\partial^4_a\phi=0$.\\
To see that the system is first class, we perform the 3+1 decomposition to the 
modified Lagrangian as well. Using the notations already introduced we obtain 
\begin{eqnarray}
H_m=\int_{\Sigma}(N\mathcal{H}+N^a\mathcal{H}_a+A_0^i G_i+A_0\underline{G})\\
\mathcal{H}=\frac{1}{\kappa\sqrt{q}}tr(2[K_a,K_b]-F_{ab})[E_a,E_b]+\frac{q_{ab}}{2\sqrt{q}}(\underline{E}^a\underline{E}^b+\underline{B}^a\underline{B}^b)+\nonumber\\
+\frac{\pi^2}{2\sqrt{q}m^2}+\frac{\sqrt{q}m^2}{2}q^{ab}(\underline{A}_a+\partial_a\phi)(\underline{A}_b+\partial_b\phi)\label{sympP1}\\
\mathcal{H}_a=F_{ab}^j E^b_j+\epsilon_{abc}\underline{E}^b\underline{B}^c+(\underline{A_a}+\partial_a\phi)\pi\\
\underline{G}=\mathcal{D}_a\underline{E}^a-\pi\label{gauge}\\
G_i=\mathcal{D}_a E^a_i\label{sympP2},
\end{eqnarray}
where $\pi$ is the conjugate momenta for $\phi$:
\begin{eqnarray}
\pi=\frac{\delta S}{\delta\dot{\phi}}=\frac{\sqrt{q}m^2(A_0-N^a A_a+\mathcal{L}_t\phi-N^a\partial_a\phi)}{N}
\end{eqnarray}
We also have the primary constraints $\Pi_N=\Pi_a=\Pi^i=\Pi_0=0$ which are the same as in the 
previous case. If we calculate $\dot{\Pi}_N$ etc., we find that $\mathcal{H},\mathcal{H}_a,G_b,\underline{G}$ 
are secondary constraints in the theory. These constraints are referred to as the  
Hamiltonian, the diffeomorphism (modulo gauge transformations), the gravitational Gauss and Maxwell Gauss-constraints. \\
The phase space carries a symplectic structure where the (nontrivial) Poisson-brackets are:
\begin{eqnarray}
\{A_a^i(x),E_j^b(y)\}=\delta_a^b\delta_j^i\delta(x-y)\\
\{\underline{A}_a(x),\underline{E}^b(y)\}=\delta^b_a\delta(x-y)\label{pb1}\\
\{\pi(x),\phi(y)\}=\delta(x-y)\label{pb2}
\end{eqnarray}
Before we turn to the quantization, it is worth to observe some of the properties 
of the above system:
\begin{itemize}
\item[-] It is easy to check that the above system is first class, i.e. the 
constraint algebra is closed. This is due to the fact that the canonical 
momenta of the scalar field appears in the Gauss constraint.
\item[-] Note that the mass only appears in the scalar constraint, which means 
that gauge- and diffeomorphism symmetries are independent of m.
\item[-] The Hamiltonian is a linear combination of constraints, which is not
true for the case where there is no scalar field, since there the Hamiltonian is 
quadratic in the Lagrange-multipliers.
\item[-] The scalar field and the Yang-Mills field is only coupled to each other 
in the scalar constraint and only through a derivative term and no scalar 
mass-term required, which means that if we will quantize this system the scalar 
field will have a totally different role than the one introduced via symmetry 
breaking.
\item[-] The term $\underline{A}_a+\partial_a\phi$ is gauge-invariant with 
respect to the gauge transformations generated by \eref{gauge}, so this is 
different to the case when we couple a scalar field to a gauge field via 
covariant derivatives (actually its more like an affine field).
\item[-] One cannot replace $m=0$ in the Hamiltonian formalism to obtain the 
usual Maxwell-field. This is not unfamiliar in loop quantum gravity, since 
it resembles to the case of the Immirzi parameter (by this analogy we do not mean any deeper connection, though). 
There the connection 
and the electric field can be rescaled as $A_a^i\to A_a^i+\beta K_a^i,E_a^i\to \frac{E_a^i}{\beta}$. 
This is a canonical transformation since the Poisson-brackets are invariant under 
this transformation. If we substitute the new quantities in the Hamiltonian, we 
find that the Gauss and diffeomorphism constraints are unchanged, but one part 
of the Hamiltonian constraint will have a term proportional with $\beta^2+1$. (see 
eg. \cite{revt}, \cite{reval} and references therein). Now consider the following 
canonical transformation:
\begin{eqnarray}
\pi\to m\pi\quad\phi\to\frac{\phi}{m}\nonumber\\
\underline{E}_a\to m\underline{E}_a\quad\underline{A}_a\to\frac{\underline{A}_a}{m}\nonumber
\end{eqnarray}
This will remove the m parameter from the mass term and furthermore this parameter 
will appear only in the Hamiltonian constraint, the other constraints will be 
independent of m.  
\end{itemize}

\section{Quantization}

In this section we will summarize the tools necessary to quantize the Proca-
field in Loop Quantum Gravity. The main advantage of the theory is that it gives  
a covariant, non-perturbative method to quantize any physical system with first 
class constraint algebra. Since the symplectically embedded Proca-field is of 
this type, we can directly apply the results achieved in Loop Quantum Gravity. 
From these the most important for us will be the quantization of general gauge 
systems (\cite{qsd}-\cite{quantdiff}) and the scalar field (\cite{fermhiggs},\cite{scalar1}). 
In the next sections we will give a brief summary on how one can apply these 
results to the present case, and we refer the reader to the above articles for 
details and proofs. 

\subsection{Quantization of a gauge field} 

Let us consider a Yang-Mills gauge field with a compact gauge group G. The Hilbert-space 
can be constructed in the following way: let $\gamma$ be an oriented graph in $\Sigma$ 
with $e_1,\dots,e_E$ edges and $v_1,\dots,v_V$ vertices. Let $h_{e_i}$ be the 
holonomy of the G-valued connection of the field evaluated along the $e_i$ edge. 
Let us define a cylindrical function with respect to a $\gamma$ graph in the following 
way:
\begin{eqnarray}
f_{\gamma}(A):=f(h_{e_1},\dots,h_{e_E})
\end{eqnarray}
where $f_{\gamma}$ is a complex valued function mapping from $G^E$. Since the 
gauge-group G is compact, there is a natural measure, the Ashtekar-Lewandowski 
measure, which enables one to have a inner product on the set of cylindrical 
functions. With this the Hilbert-space of the Yang-Mills field is defined as
the set of all cylindrical functions which are square-integrable with respect 
to the above measure: 
\begin{eqnarray}
\mathcal{H}:=L_2(\bar{\mathcal{A}},d\mu_{AL,G})
\end{eqnarray}
In our case, $G=SU(2)\times U(1)$, so 
\begin{eqnarray}
\mathcal{H}_{G,YM}:=L_2(\bar{\mathcal{A}}_{SU(2)},d\mu_{SU(2)})\otimes L_2(\bar{\underline{\mathcal{A}}}_{U(1)},d\mu_{U(1)})\label{HilbertG}
\end{eqnarray}
In order to analyze the action of the Hamiltonian and to compute its kernel,
it is convenient to introduce a complete orthonormal basis on the Hilbert-space
\eref{HilbertG}.\\
On the space of $L_2(\bar{\mathcal{A}}_{SU(2)},d\mu_{SU(2)})$ these are called 
\textit{spin network functions} and defined as follows: let $\gamma\in\Sigma$ 
be a graph and denote its edges and vertices respectively by $(e_1,\dots,\e_N)$ 
and $(v_1,\dots,v_V)$. Associate a coloring to each edge defined by a set of 
irreducible representations $(j_1,\dots,j_N)$ of SU(2) (half-integers) and 
contractors $(\rho_1,\dots,\rho_V)$ to the vertices where $\rho_l$ is an 
intertwiner which maps from the tensor product of representations of the 
incoming edges at the vertex $v_l$ to the tensor product of representations 
of the outgoing edges. A spin network state is then defined as
\begin{eqnarray}
|T(A)>_{\gamma,\vec{j},\vec{\rho}}:=\bigotimes_{i=1}^N j_i(h_{e_i}(A))\cdot\bigotimes_{k=1}^V\rho_k\label{spin1}
\end{eqnarray}
where $\cdot$ stands for contracting at each vertex $v_k$ the upper indices of the
matrices corresponding to all the incoming edges and the lower indices of the 
matrices assigned to the outgoing edges with all the indices of $\rho_k$.\\
In the case of $L_2(\bar{\mathcal{A}}_{U(1)},d\mu_{U(1)})$ one must simply replace 
SU(2) with U(1) in the above definition - these are called \textit{flux network
functions} (\cite{mq2}). Since U(1) is a commutative group, we will have the following definition:
for each edge $e_i$ of the graph associate an integer $n_i$. Then the flux network 
function is defined as 
\begin{eqnarray}
|F(\underline{A})>_{\gamma,\vec{l}}:=\prod_{i=1}^N (h_{e_i}(\underline{A}))^{n_i}
\end{eqnarray}   
What remains is to define the operators corresponding to the connection and the 
electric field on the Hilbert-space. If we want to implement the Poisson-brackets 
in the quantum theory in a diffeomorphism covariant way, we have to use smeared 
versions of these fields. In the case of gauge fields the natural candidates are 
the holonomy and the electric flux respectively:
\begin{eqnarray}
h_e(A)=\mathcal{P}\exp{\int_e A}\label{sm1}\\
E(S)=\int_S*E\label{sm2},
\end{eqnarray}
where $e$ is a path and $S$ is a surface in $\Sigma$. Then the action of the 
corresponding operators will be defined via the following way:
\begin{eqnarray}
\hat{h}_e(A)f(A):=h_e(A)f\\
\hat{E}(S) f(A):=i\bar{h}\{E(S),f(A)\}\label{der}
\end{eqnarray}  
Let us make the action of the electric flux operator a bit more explicit for both the 
gravitational and the Yang-Mills part. The latter case is simple, since the U(1) 
group is commutative. On a flux network function the action of this operator is 
\begin{eqnarray}
\underline{\hat{E}}(S)|F>_{\gamma,\vec{n}}=i\bar{h}\sum_{i=1}^N n_i \kappa(S,e_i)|F>_{\gamma,\vec{n}}
\end{eqnarray}
where 
\begin{displaymath}
\kappa(S,e)=\left \{ \begin{array}{ll} 0, & \textrm{if $e\cap S=0$ or $e\cap S=e$ modulo the endpoints}\\
+1, & \textrm{if e lies above S}\\
-1, & \textrm{if e lies below S}\\
\end{array} \right.
\end{displaymath}
In the case of the gravitational part introduce the left and right invariant 
vector fields on SU(2):
\begin{eqnarray}
R_h f(g)=\frac{d}{dt}f(e^{th}g)\textrm{ and } L_h f(g)=\frac{d}{dt}f(ge^{th})\nonumber
\end{eqnarray}
Using these quantities the action of the flux operator on a cylindrical function is
\begin{eqnarray}
\hat{E}(S) f_{\gamma}(A)=\frac{1}{4}\sum_e\kappa(S,e)(\delta_{e\cap S,b(e)}R^{(e)}_{\tau_i}+\delta_{e\cap S,f(e)}L^{(e)}_{\tau_i})f_{\gamma}(A)
\end{eqnarray}
where b(e) and f(e) is the beginning- and endpoints of the edge e and $R^{(e)}$ 
is $R$ on the copy of SU(2) labelled by e.
\subsection{Quantization of the scalar field}
Though the scalar field requires more careful treatment, the results are very 
similar to the Yang-Mills case. The main difference is that in this case one 
has to use so-called point holonomies of the form
\begin{eqnarray}
U(v)=\exp{(i\phi(x_v))}
\end{eqnarray}
Then a cylindrical function with respect to a $\gamma$ graph has the form
\begin{eqnarray}
f_{\gamma}(U):=f(U_{v_1},\dots,U_{v_V})\label{scal1}
\end{eqnarray}
where $v_1,\dots,v_V$ are the vertices of the graph. Then the Hilbert space will be
\begin{eqnarray}
\mathcal{H}_U:=L_2(\bar{\mathcal{U}},d\mu_{\phi})
\end{eqnarray}

Recently it was shown that a more general description should be used because in the 
original approach the configuration variables are periodic functions, which do not 
suffice to separate the points of configuration space (for details see \cite{scalar1} and \cite{scalar2}).
So instead of $U(v)$ we define the configuration variable to be
\begin{eqnarray}
U(\lambda,v)=\exp{(i\lambda\phi(x_v))}.
\end{eqnarray}
Given a graph $\gamma$ with vertices $(v_1,\dots,v_N)$ and real numbers $\lambda_1,\dots,\lambda_N$ 
associated to the vertices, let $Cyl_N$ be the set of finite linear combinations of the following 
functions:
\begin{eqnarray}
|D(U)>_{\gamma,\vec{\lambda}}:=\prod_{i=1}^{N}(U(\lambda_i,v_i))
\end{eqnarray}
Then the completion of $Cyl=\cup_N Cyl_N$ will be the $C^*$ algebra of configuration observables. 

Because the point-holonomy is not smeared, the regulated version of the 
canonical momenta $\pi$ is smeared in three dimensions:
\begin{eqnarray}
P_B=\int_B\pi\label{scal2},
\end{eqnarray}
where $B$ is an open ball in $\Sigma$.\\
The operator corresponding to \eref{scal1} is defined the same way as was done 
in the case of the Yang-Mills field, but the operator version of \eref{scal2} 
needs careful treatment, because the functional derivative of $U(v)$ with 
respect to $\phi$ is meaningless. The precise calculations can be found in 
\cite{fermhiggs}, the result is
\begin{eqnarray}
\hat{P}_B|D(U)>_{\gamma,\vec{\lambda}}:=-i\bar{h}\chi_B(v)X(v)|D(U)>_{\gamma,\vec{\lambda}},
\end{eqnarray}
where $\chi_B(v)$ is zero unless $v\in B$ and $X(v)$ is the symmetric sum of 
left and right invariant vector fields at $U(v)$. But since we have a Abelian group, we can 
write that $X(v)=X_L(v)=X_R(v)$, so we obtain
\begin{eqnarray}
\hat{P}_B|D(U)>_{\gamma,\vec{\lambda}}:=-i\bar{h}\sum_{j=1}^N\chi_B(v_j)\lambda_j|D(U)>_{\gamma,\vec{\lambda}}
\end{eqnarray}
\\ \\
Now let us apply these results to the case of the Proca-field. Since we have an 
$SU(2)\times U(1)$ Yang-Mills field and a real scalar field, the Hilbert-space 
will be
\begin{eqnarray}
\mathcal{H}=L_2(\bar{\mathcal{A}}_{SU(2)},d\mu_{SU(2)})\otimes L_2(\bar{\mathcal{A}}_{U(1)},d\mu_{U(1)})\otimes L_2(\bar{\mathcal{U}},d\mu_{U(1)})
\end{eqnarray}
with basis (later referred to as generalized spin network functions)
\begin{eqnarray}
|S>_{\gamma,\vec{j},\vec{\rho},\vec{l},\vec{\lambda}}=|T(A)>_{\gamma,\vec{j},\vec{\rho}}\otimes|F(\underline{A})>_{\gamma,\vec{l}}\otimes|D(U)>_{\gamma,\vec{\lambda}}\label{genspin}
\end{eqnarray}
{\it Remark}: We have seen that in the original (not symplectically embedded) case \eref{origH} we arrived to 
a second class constraint algebra, thus we had  to introduce Dirac-brackets instead of Poisson-brackets. 
Because of this, the above definitions of the momentum operators should be modified by replacing the 
Poisson-brackets with Dirac-brackets. Specifically the momentum operators should be redefined in the following way:
\begin{eqnarray}
\hspace{-2cm}\hat{E}(S)_D f(A):=i\bar{h}\{E(S),f(A)\}_D=\nonumber\\
\hspace{-2cm}=i\bar{h}\{E(S),f(A)\}-i\bar{h}\int d^3 x d^3 y \{E(S),C_i(x)\}(M^{(-1)})_{ij}(x,y)\{C_j(y),f(A)\},
\end{eqnarray}
where $C_i$ are the second class constraints $\bar{\underline{G}}$ and $\bar{H}$. 
This does not modify the properties of the momentum operators, 
since the Dirac-brackets have the same properties as the Poisson-brackets. Only the action on spin 
network functions changes, which becomes more complicated since $\tilde{M}(x,y)$ is not explicit and depend also on 
$\underline{A}_a$ and $\underline{E}_a$. 
In fact only the momentum operator of the Maxwell-field changes, since $\{\tilde{G},A_a^i\}=\{\tilde{G},E_a^i\}=0$.  
The only question is whether the momentum operator defined above is a well defined operator on the Hilbert space, since 
it is not trivial if its action is a cylindrical function. The more detailed analysis of this question can be found in 
section 6. 

\section{The Hamiltonian of the Proca-field}

The total (non-smeared) Hamiltonian of the symplectically embedded Proca-field has the form
\begin{eqnarray}
H=\int_{\Sigma}N(\mathcal{H}_{GE}+\mathcal{H}_{GL}+\mathcal{H}_{E}+\mathcal{H}_{B}+\mathcal{H}_{P}+\mathcal{H}_{M})\label{hamilton}\\
\mathcal{H}_{GE}=\frac{2}{\kappa}\epsilon^{abc}tr(F_{ab}\{A_c,V\})\nonumber\\
\mathcal{H}_{GL}=\frac{8}{\kappa^3}\epsilon^{abc}tr(\{A_a,K\}\{A_b,K\}\{A_c,V\})\nonumber\\
\mathcal{H}_{E}=\frac{q_{ab}}{2\sqrt{q}}\underline{E}^a\underline{E}^b\nonumber\\
\mathcal{H}_{B}=\frac{q_{ab}}{2\sqrt{q}}\underline{B}^a\underline{B}^b\nonumber\\
\mathcal{H}_{P}=\frac{\pi^2}{2\sqrt{q}m^2}\nonumber\\
\mathcal{H}_{M}=\frac{\sqrt{q}m^2}{2}q^{ab}(\underline{A}_a+\partial_a\phi)(\underline{A}_b+\partial_b\phi),\nonumber
\end{eqnarray}
where
$H_{GE}$ and $H_{GL}$ are the Euclidean and Lorentzian part of the gravitational, 
$H_{E}$ and $H_{B}$ are the electric and magnetic part of the Yang-Mills Hamiltonian, 
$H_P$ is the kinetic term and $H_{M}$ is the mass term (in the following we consider 
only the symplectically embedded Hamiltonian).
To arrive at a well defined, diffeomorphism covariant Hamiltonian operator, one
first has to get rid of the $\sqrt{q}$ quantity from the denominator. This is 
achieved via using the identity (introduced by Thiemann):
\begin{eqnarray}
\frac{1}{\kappa}\{A^i_a,V\}=2sgn(det\ e)e^i_a
\end{eqnarray}
The next step is to rewrite the Hamiltonian in terms of holonomies and fluxes.
To do this, one introduces a regularisation scheme, a $\Delta$ triangulation of 
$\Sigma$. Each tetrahedra in $\Delta$ has a base-point $v$ with three outgoing 
segments $s_i\ i=1,2,3$. Denote the arcs connecting the endpoints of $s_i$ and 
$s_j$ by $a_{ij}$, so we can form the loop $\alpha_{ij}:=s_i\circ a_{ij}\circ s_j^{-1}$. 
After an appropriate point-splitting, one approximates the quantities in the 
Hamiltonian by their smeared counterparts, substitutes the corresponding 
operators and takes the limit. This can be done with each of the terms
appearing in \eref{hamilton}.
The details of the computations are quite lengthy so here we will only present the 
regulated Hamiltonians and refer the reader to \cite{qsd5} or \cite{mq1}
where one can find the regularisation of all terms except the mass term. Since the latter 
is important in our analysis we will sketch the derivation of that particular term.
The regulated Hamiltonians are 
\begin{eqnarray}
\hat{H}_{GE}=-\frac{8}{3i\bar{h}\kappa}\sum_v\frac{N(v)}{E(v)}\epsilon^{ijk}tr(h_{\alpha_{ij}}h_{s_k}[h^{-1}_{s_k},\hat{V}])\\
\hat{H}_{GL}=\frac{64}{3i\bar{h}^3\kappa^3}\sum_v\frac{N(v)}{E(v)}\epsilon^{ijk}tr(h_{s_i}[h^{-1}_{s_i},K_v]h_{s_j}[h^{-1}_{s_j},K_v]h_{s_k}[h^{-1}_{s_k},V_v])\\
\hat{H}_{YM}=-\frac{32}{9\bar{h}^2\kappa^2}\sum_v\frac{N(v)}{E(v)^2}\sum_{v(\Delta)=v(\Delta`)=v}\epsilon^{JKL}\epsilon^{MNP}\times\nonumber\\
\times\hat{Q}^i_{s_L(\Delta)}(v,\frac{1}{2})\hat{Q}^i_{s_P(\Delta`)}(v,\frac{1}{2})\times\nonumber\\
\times\big[\underline{\hat{h}}_{\alpha_{JK}}\underline{\hat{h}}_{\alpha`_{MN}}-\underline{\hat{E}}(F_{JK})\underline{\hat{E}}(F`_{MN})\big]\\
\hat{H}_P=\frac{8}{81m^2\bar{h}^4\kappa^6}\sum_v N(v)\Big(\frac{X(v)}{E(v)}\Big)^2\sum_{v(\Delta)=v(\Delta`)=v}\epsilon^{IJK}\epsilon^{LMN}\epsilon_{ijk}\epsilon_{lmn}\times\nonumber\\
\times\hat{Q}^i_{s_I(\Delta)}(v,\frac{1}{2})\hat{Q}^j_{s_J(\Delta)}(v,\frac{1}{2})\hat{Q}^k_{s_K(\Delta)}(v,\frac{1}{2})\times\nonumber\\
\times\hat{Q}^l_{s_L(\Delta`)}(v,\frac{1}{2})\hat{Q}^m_{s_M(\Delta`)}(v,\frac{1}{2})\hat{Q}^n_{s_N(\Delta`)}(v,\frac{1}{2}),
\end{eqnarray}
where\\
$\hat{Q}^k_{e}(v,r)=tr(\tau_k h_e[h^{-1}_e,V(v)^r])$\\
$E(v)=\frac{n(n-1)(n-2)}{6}$, where n is the valance of the vertex v\\
$F_{JK}$ is a surface parallel to the face determined by $s_J$ and $s_K$\\
What remains is the quantization of the mass term. It is easy to see that this term 
can be obtained from the derivative term of \cite{qsd5} with the replacement 
$\partial\phi\to\partial\phi+A$, so the calculation is completely similar. 
From $q^{ab}\sqrt{q}=\frac{E_i^a E_i^b}{\sqrt{q}}$ and $E_i^a=\epsilon^{acd}\epsilon_{ijk}\frac{e_c^j e_d^k}{2}$ we have
\begin{eqnarray}
H_M=\frac{m^2}{2}\int d^3 x\int d^3 y N(x)\chi_{\epsilon}(x,y)\epsilon^{ijk}\epsilon^{ilm}\epsilon_{abc}\epsilon_{bef}\times\nonumber\\
\times\frac{((\partial_a\phi+\underline{A}_a) e_b^j e_c^k)(x)}{\sqrt{V(x,\epsilon)}}\frac{((\partial_b\phi+\underline{A}_b) e_e^m e_f^n)(y)}{\sqrt{V(y,\epsilon)}}=\nonumber\\
=\frac{m^2}{2}\Big(\frac{2}{3\kappa}\Big)^4\int N(x)\epsilon^{ijk}(\partial\phi+\underline{A})(x)\wedge\{A^j(x),V(x,\epsilon)^{3/4}\}\wedge\{A^k(x),V(x,\epsilon)^{3/4}\}\times\nonumber\\
\times\int\chi_{\epsilon}(x,y) \epsilon^{imn}(\partial\phi+\underline{A})(y)\wedge\{A^m(y),V(y,\epsilon)^{3/4}\}\wedge\{A^n(y),V(y,\epsilon)^{3/4}\}\label{reg1}
\end{eqnarray}
Now we shell introduce the familiar triangulation (with $v=s(0)$) and use the 
following (exploiting the fact that we are dealing with Abelian fields):
\begin{eqnarray}
U(1,s(\delta t))=\exp[i(\phi(v)+\delta t\dot{s}^a\partial_a\phi(v)+o(\delta t^2))]\nonumber\\
\underline{h}_s(0,\delta t)=\exp[i(\delta t\dot{s}^a \underline{A}_a+o(\delta t^2))]\label{regskem}
\end{eqnarray}
With this we have that
\begin{eqnarray}
U(1,s(\delta t))\underline{h}_s(0,\delta t)U(1,v)^{-1}-1=i\delta t\dot{s}^a(\partial_a\phi(v)+\underline{A}_a(v))+o(\delta t^2),
\end{eqnarray}
so we have
\begin{eqnarray}
\int_{\Delta}(\partial\phi+\underline{A})(x)\wedge\{A^j(x),V(x,\epsilon)^{3/4}\}\wedge\{A^k(x),V(x,\epsilon)^{3/4}\}\approx\nonumber\\
\approx\frac{2}{3}\epsilon^{mnp}[U(1,s_m(\Delta))\underline{h}_{s_m(\Delta)}U(1,v(\Delta))^{-1}-1]\hat{Q}^j_{s_n(\Delta)}(v,\frac{3}{4})\hat{Q}^k_{s_p(\Delta)}(v,\frac{3}{4})
\end{eqnarray}
Substituting this into \eref{reg1} and taking the limit $\epsilon\to 0$ we have
\begin{eqnarray}
\hat{H}_M=\frac{m^2}{2\bar{h}^4\kappa^4}\Big(\frac{4}{3}\Big)^6\sum_v\frac{N(v)}{E(v)^2}\sum_{v(\Delta)=v(\Delta`)=v}\epsilon^{ijk}\epsilon^{ilm}\epsilon_{npq}\epsilon_{rst}\times\nonumber\\
\times[U(1,s_n(\Delta))\underline{h}_{s_n(\Delta)}U(1,v)^{-1}-1][U(1,s_r(\Delta`))\underline{h}_{s_r(\Delta`)}U(1,v)^{-1}-1]\times\nonumber\\
\times\hat{Q}^j_{s_p(\Delta)}(v,\frac{3}{4})\hat{Q}^k_{s_q(\Delta)}(v,\frac{3}{4})\hat{Q}^l_{s_s(\Delta`)}(v,\frac{3}{4})\hat{Q}^m_{s_t(\Delta`)}(v,\frac{3}{4})
\end{eqnarray}
This part of the Hamiltonian looks problematic due to open ends of the holonomies, but since U(1) gauge invariance 
is studied with respect to \eref{gauge} and $\{\underline{G},\underline{A}_a+\partial_a\phi\}=0$ the term $\hat{H}_M$ 
is U(1) gauge invariant (open ends of holonomies are compensated by the scalar field, since the latter is defined 
in the vertices). Further more this term is also diffeomorphism and SU(2) gauge invariant, thus during 
quantization we do not come up against any problems.\\
The total Hamiltonian of the (symplectically embedded) Proca-field is
\begin{eqnarray}
\hat{H}=\hat{H}_{GE}+\hat{H}_{GL}+\hat{H}_P+\hat{H}_M\nonumber
\end{eqnarray}
As Thiemann noticed earlier for similar systems, the Hamiltonian is well-defined, i.e. it doesn't suffer from UV 
divergences, and this is achieved not via renormalisation or spontaneous 
symmetry breaking but treating the gravitational field dynamical.

\section{Kernel of the Hamiltonian of the Proca-field}

\subsection{Complete solution}

Though this Hamiltonian is quite complicated, there are a lot of relevant 
informations that can be extracted from it.
First let's look at the action of the different terms in the Hamiltonian on a 
generalized spin network state.\\ 
The action of the gravitational term $\hat{H}_G$ changes the graph (as pointed out in \cite{revt}) in a way 
that it adds additional edges (specifically extraordinary edges) to the graph and 
changes the intertwiners, but it doesn't affect the labels which correspond to 
the matter fields. The other parts of the 
Hamiltonian describe the matter fields. Their structure is similar: they all 
contain matter operators and the $\hat{Q}$ operator in some way which encode 
the interaction of the fields with gravity. This operator only changes the 
intertwiners, it doesn't change neither the colorings or the graph itself. The 
derivative operators - the electric part of the Yang-Mills and the kinetic term -, 
don't change the graphs, only the coefficients, but the mass term and the 
magnetic term does.\\
It is not an obvious question whether this Hamiltonian possesses a non-trivial 
kernel, but we will show that the construction of generating a solution to the 
Hamiltonian constraint, which was introduced by Thiemann, can be generalized to 
the present case.\\
Let $|T>_{\gamma,\vec{\rho},\vec{l},\vec{m},\vec{n}}:=|T>_s$ be a 
spin-color network state. Then $<\Phi|$ is in the kernel of the Hamiltonian 
of the Proca-field if for all $|T>_s$ we have
\begin{eqnarray}
<\Phi|\hat{H}|T>_s=0
\end{eqnarray}
The key observation of Thiemann was that the Hamiltonian of gravity acts as it 
generates so called extraordinary edges (see details in \cite{revt} or \cite{reval}). 
and with this the kernel can be constructed in the following way: Denote the set of 
labelled graphs (spin-nets) $S_0\in (\gamma_0,l_0)$ which contain no extraordinary edges 
(these are the ``sources''). Then compute $S_{n+1}$ by acting $\hat{H}_G$ on the 
elements of $S_n$ and decomposing them into spin-network states. The main advantage 
of the sets $S_n$ that 1.) they are disjoint, i.e. $S_n\cap S_m=\delta_{mn}$ and 2.) 
finding a general diffeomorphism invariant solution to the Hamiltonian-constraint reduces 
to finding a solution on a finite subspace.\\
Since we are only interested in solutions which are diffeomorphism invariant, we use $T_{[s]}$ 
instead of $T_s$, where [s] labels the diffeomorphism invariant distribution.
In particular, let the ansatz for a solution be of the form
\begin{eqnarray}
<\Psi|:=\sum_{i=1}^N \sum_{[s]\in [S^{n_i}]} c_{[s]}<T|_{[s]}\nonumber
\end{eqnarray}
Since the Euclidean part of the gravitational Hamiltonian maps from $S^{(n)}$ 
to $S^{n+1}$, we have that the condition
\begin{eqnarray}
\sum_{i=1}^N \sum_{[s]\in S^{n_i}} c_{[s]}<T|_{[s]}\hat{H}_{GE}|T>_{[s']}=0
\end{eqnarray}
is non-trivial if and only if $[s']\in [S^{n_i-1}]$. Since $\hat{H}_{GE}=\sum_vN(v)\hat{H}_{GE}(v)$ 
and the above equation has to hold for all possible N, we have
\begin{eqnarray}
\sum_{i=1}^N \sum_{[s]\in S^{n_i}} c_{[s]}<T|_{[s]}\hat{H}_{GE}(v)|T>_{[s']}=0
\end{eqnarray}
for each choice of finite number of vertices v and spin nets. Thus, we arrived at 
a finite system of liner equations with finite number of coefficients.\\
In the case of the Proca-field we have three fields. Since the orthonormal base 
is of the form $|T>\otimes|F>\otimes|D>$, we will first look 
for analogs of the sets $S^{(n)}$ in the case of the scalar-field and the Yang-Mills 
field.\\
The case of the scalar field is simple: denote the set $S^{(0)}(U)(\gamma)$ of 
all colored graphs, that all vertices are labelled by zero. Now define $S^{(n+1)}(U)$ 
by acting with $U(1,v)$ (for all possible v) on every element of $S^{(n)}(U)$. 
From the simple action of $U(1,v)$ it is clear that this is equivalent to that 
the elements of $S^{(n)}(U)$ are those colored graphs, for which the sum of the 
vertex colorings are n. If we look at the form of $\hat{H}_M$ we find that 
it maps from $S^{(n)}(U)$ to $S^{(n)}(U)\cup S^{(n+1)}(U)\cup S^{(n+2)}(U)$.\\
The case of the Yang-Mills field is a bit more complicated, because both $\hat{H}_B$ 
and $\hat{H}_M$ changes the graph. The former adds two (Yang-Mills) loops with 
color 1, while the latter increases the color of two edges by one 
(non-existent edges can be treated like they were edges with coloring zero). 
This is why in this case the analogs of $S^{(n)}$ will have two indices. 
Denote by $S^{(n,m)}(YM)$ the set of labelled graphs which have n loops with 
color 1 and the sum of the colors on all edges are m. It is easy to see that 
these sets are disjoint and the action of the $\hat{H}_B$ and $\hat{H}_M$ operators 
are the following: while $\hat{H}_B$ maps from $S^{(n,m)}(YM)$ to $S^{(n+2,m+2)}(YM)$, 
$\hat{H}_M$ maps from $S^{(n,m)}(YM)$ to $S^{(n,m+2)}(YM)\cup S^{(n,m+1)}(YM)\cup S^{(n,m)}(YM)$. 
It follows from the 
construction that $S^{(0,0)}(YM)$ will contain labelled graphs with zero colorings 
on all edges. Also note that the set $S^{(n,m)}(YM)$ is empty unless $n\geq m$.\\
Now, let the ansatz for the solution to the kernel of the Proca-field be of the 
form
\begin{eqnarray}
\hspace{-2cm}<\Psi|:=\sum_{i,j,k,l}\sum_{[s]\in S^{(n_i)}}\sum_{[f]\in S^{(m_j,p_k)}(YM)}\sum_{[d]\in S^{(q_l)}(U)}c_{[s],[f],[d]}<T|_{[s]}\otimes<F|_{[f]}\otimes<D|_{[d]}\label{solution}
\end{eqnarray}
Now \eref{solution} is in the kernel of the Hamiltonian of the Proca-field, if 
for all [s]', [f]', [d]'
\begin{eqnarray}
<\Psi|\hat{H}|T>_{[s]'}\otimes|F>_{[f]'}\otimes|D>_{[d]'}=0\label{solution2}
\end{eqnarray}
With the same reasoning as before this condition is non-trivial if\\
-$[s]'\in [S^{n_i-1}]\cup [S^{n_i-2}]$ (the union of the two sets is necessary if one takes 
the action of $\hat{H}_{GL}$ also into account, since this latter operator adds two extraordinary 
edges)\\
-$[d]'\in [S^{(q_l)}(U)]\cup [S^{(q_l-1)}(U)]\cup [S^{(q_l-2)}(U)]$ \\
-$[f]'\in [S^{(m_j-2,p_k-2)}(YM)]\cup [S^{(m_j,p_k-2)}(YM)]\cup [S^{(m_j,p_k-1)}(YM)]\cup [S^{(m_j,p_k)}(YM)]$

\subsection{Special solutions}

Since the system of equation \eref{solution2} is very complicated, it is useful 
to take some special solutions in order to understand the full theory. Let 
$<0|_{YM}$ be the "vacuum'' flux network state of the Yang-Mills sector, which 
means that it has no Yang-Mills colors on either edge (note that this is not actually 
the familiar vacuum state as was shown in \cite{mq2}, since we are not dealing with Fock-spaces). 
Similarly, denote the 
vacuum dust network state and vacuum spin network state by $<0|_U$ and $<0|_G$ respectively. 
Now it is easy to check that the state 
$<\Psi|_G\otimes<0|_{YM}\otimes0|_U$ is a solution of \eref{solution2} if 
$<\Psi_G|$ is the solution of the gravitational part of the  Hamiltonian (this 
is because these "vacuum states" are annihilated by the corresponding derivative 
operators in $\hat{H}_P$ and $\hat{H}_{E_{YM}}$, and are orthogonal to every
state created by the operators in $\hat{H}_{YM}$ and $\hat{H}_U$). So these 
special states can be interpreted as pure gravity.\\
Now let us check states of the form $<0|_G\otimes<\Psi|_{YM}\otimes<0|_U$. 
It is easy to show that this is in the kernel of the Hamiltonian for all $<\Psi|_{YM}$!
In fact the same is true for states of the form $<0|_G\otimes<\Psi|_{YM}\otimes<\Psi|_U$. 
These states are obviously nonphysical since the expectation value of the volume, area and 
length operators of these states are all zero for all volumes, surfaces and curves respectively.
(It is worth mentioning that these states are in the kernel of 
all Hamiltonian which have density weight one and composed only from the gravitational, 
Yang-Mills and scalar fields)\\
Let us check whether there are solutions of the form $<\Psi|_G\otimes<\Psi|_{YM}\otimes<0|_U$, 
where $<\Psi|_G\otimes<\Psi|_{YM}$ is in the kernel of $\hat{H}_G+\hat{H}_{YM}$. The answer is 
yes, if $<\Psi|_G\otimes<\Psi|_{YM}$ contains only flux networks that have U(1) colors only on 
the loops, not on the edges, since in 
this case these states are annihilated by the operator $\hat{H}_M$. These states are in the 
subset of the kernel of the Yang-Mills field coupled to gravity. Since currently we do not have 
a semi-classical description of the above system it will be for future investigations to check 
the physical meaning of these states. But if we look at the limit $m\to 0$, we find that these 
states will be solutions, since $<\Psi|_G\otimes<\Psi|_{YM}\otimes<0|_U X(v)|\phi>=0$ for all $|\phi>$ and 
\begin{eqnarray}
<\Psi|_G\otimes<\Psi|_{YM}\otimes<0|_U(\hat{H}-\hat{H}_G-\hat{H}_{YM})|\phi>=\nonumber\\
=<\Psi|_G\otimes<\Psi|_{YM}\otimes<0|_U\hat{H}_M|\phi>\to 0
\end{eqnarray}

\section{Gauge fixing}

We used the symplectically embedded Proca-field to avoid implementing the Dirac-brackets in 
the quantum theory. This approach led to a well defined quantum theory as it was shown in the previous 
sections. This was achieved by introducing an auxiliary scalar field
to the formalism. The theory we gained is equivalent to the original one since if the constraints are 
solved the scalar field disappears automatically. But this equivalence is not manifest if we do 
not solve the constraints, thus we need to introduce gauge fixing. As we shall see this will again lead 
to a system with second class constraints like in the original theory (without the scalar field). Below 
we outline how the quantization of such systems could be handled.\\
The key observation is, as we pointed this out in section 2, that the original Lagrangian can be obtained 
via the gauge fixing $\mathcal{D}^4_a\phi=0$. The strategy will be to implement this condition 
in the Hamiltonian formalism. 
If we compare the original Hamiltonian with the Hamiltonian of the symplectically embedded Proca-field we find that 
if we substitute
\begin{eqnarray}
\partial_a\phi=0\\
\pi=\sqrt{q}m^2\frac{A_0-N^a\underline{A}_a}{N}
\end{eqnarray}
into \eref{sympP1}-\eref{gauge}, we obtain the constraints \eref{co1}-\eref{co3}. So if we introduce 
the two extra conditions (constraints)
\begin{eqnarray}
C_a=\partial_a\phi=0\label{gf1}\\
C=\pi-\sqrt{q}m^2\frac{A_0-N^a\underline{A}_a}{N}=0\label{gf2},
\end{eqnarray}
we arrive to the original case.
One may ask how come the original theory have second class constraints while the symlectically 
embedded one has only first class constraints. The trick is that, as it was pointed out in \cite{gfixing}, 
the conditions \eref{gf1} and \eref{gf2} are also constraints and if we include these to the 
constraint algebra, we obtain a system with second class constraints.\\
(In the case of second class constraints one must use Dirac-brackets, so it is natural to ask what did 
we gain with the symplectic embedding? Actually, the main advantage of this method that we were able to quantize 
the theory without introducing the Dirac-brackets. The Dirac-brackets are only needed when one fixes the gauge.)\\
Now what remains is to calculate the Dirac-bracket and implement the two conditions \eref{gf1} and \eref{gf2} in the quantum theory. 
First we need the Poisson brackets of the new constraints with the existing ones. If we define 
the same linear combinations \eref{comb1},\eref{comb2},\eref{comb3} as for the original case, we 
find that these are first class constraints. Also with the same reasoning as we did there one finds 
that $G_i$ and $\mathcal{H}_a+G_i A^i_a+\underline{G}\underline{A}_a$ are also first class constraints. 
Thus we have four second class constraints: $\mathcal{H},\underline{G},C_a,C$. The elements of the 
antisymmetric matrix $M^{(P)}_{ij}(x,y)$ are therefor the following:
\begin{eqnarray}
M^{(P)}_{12}=\{\mathcal{H}(x),\underline{G}(y)\}=0\\
M^{(P)}_{13}=\{\mathcal{H}(x),C_a(y)\}=\frac{\pi(x)}{m^2\sqrt{q}(x)}\mathcal{D}^{(y)}_a\delta(x-y)\\
M^{(P)}_{14}=\{\mathcal{H}(x),C(y)\}=-m^2\sqrt{q}(x)(\underline{A}_a+\partial_a\phi)\mathcal{D}^{a(x)}\delta(x-y)+\nonumber\\
+m^2\frac{\sqrt{q}(y)}{\sqrt{q}(x)}\frac{N^a(y)\underline{E}_a(x)}{N(y)}\delta(x-y)\\
M^{(P)}_{23}=\{\underline{G}(x),C_a(y)\}=-\mathcal{D}^{(y)}_a\delta(x-y)\\
M^{(P)}_{24}=\{\underline{G}(x),C(y)\}=m^2\frac{\sqrt{q}(y)}{N(y)}N^a(y)\mathcal{D}^{(x)}_a\delta(x-y)\\
M^{(P)}_{34}=\{C_a(x),C(y)\}=-\mathcal{D}^{(x)}_a\delta(x-y)
\end{eqnarray} 
With the inverse matrix $(M^{(P)})^{(-1)}_{ij}(x,y)$ one can define the Dirac-brackets the 
similar way as in the first chapter, the only difference is that now we have six second 
class constraints and the matrix $M^{(P)}_{ij}(x,y)$ is much more complicated. \\
After this we quantize the theory in the following way. The Hilbert space and the configuration 
variables are defined as in the original (first class constraint) case, but we  
have to redefine the momentum operators $\hat{\underline{E}}(S)$ and $\hat{P}_B$ (see the {\it Remark} 
at the end of section 3). Now let us check how our new constraints can be interpreted in the quantum theory. 
Because of the complicated Dirac-bracket, the precise action of the operator version of \eref{gf2} is left for future studies. 
The constraint \eref{gf1} on the other hand is much more simple. If we look at the regularisation of the mass term, 
specifically at equations \eref{regskem}, we see that 
\begin{eqnarray}
U(1,s(\delta t))-U(1,v)=i\delta t\dot{s}^a\partial_a\phi(v)+o(\delta t^2),
\end{eqnarray}
so the constraint can be implemented as $<\Psi|$ is in its kernel only if 
\begin{eqnarray}
<\Psi|(U(1,b(e))-U(1,f(e)))|\psi>=0\label{cond}
\end{eqnarray}
for all $|\psi>$ and all e edge (b(e) is the beginning-, f(e) is the endpoint of the edge). 
Let $|\psi>$ be a basis element, that is 
\begin{eqnarray}
|\psi>=|S>_{\gamma,\vec{j},\vec{\rho},\vec{l},\vec{\lambda}}=|T(A)>_{\gamma,\vec{j},\vec{\rho}}\otimes|F(\underline{A})>_{\gamma,\vec{l}}\otimes|D(U)>_{\gamma,\vec{\lambda}}
\end{eqnarray}
It is obvious that the action of the operator in \eref{cond} will be
\begin{eqnarray}
(U(1,b(e))-U(1,f(e)))|\psi>=|S>_{\gamma,\vec{j},\vec{\rho},\vec{l},\vec{\lambda}_1}-|S>_{\gamma,\vec{j},\vec{\rho},\vec{l},\vec{\lambda}_2},
\end{eqnarray}
where $\lambda_1$ and $\lambda_2$ is obtained by increasing the value of $\lambda_v$ in the appropriate vertex by one. 
Since the condition is implemented in all vertices, we have that the coefficient of 
$<S|_{\gamma,\vec{j},\vec{\rho},\vec{l},\vec{\lambda}}$ in $<\Psi|$ is the same as the coefficient of 
$<S|_{\gamma,\vec{j},\vec{\rho},\vec{l},\vec{\lambda}'}$ if both $|D(U)>_{\gamma,\vec{\lambda}}$ and 
$|D(U)>_{\gamma,\vec{\lambda}'}$ are elements of $S^{(n)}(U)$ for a fixed value of n. In other words
if we substitute \eref{solution} into the above constraint we get the following condition 
on the coefficients
\begin{eqnarray}
c_{[s],[f],[d]_1}-c_{[s],[f],[d]_2}=0
\end{eqnarray}
for all $[s],[f],[d]_1,[d]_2$, where $[d]_1,[d]_2$ are in the same set $S^{(n)}(U)$. This condition has 
non-trivial solutions, for example the case where only those coefficients of $<S|_{\gamma,\vec{j},\vec{\rho},\vec{l},\vec{\lambda}}$ 
are not zero where $\lambda_v$ is the same for all vertices.\\
This was the case when one first implements gauge fixing, then quantize the system. 
One may ask whether it is possible to first quantize the system, then do gauge fixing. This is a problematic issue 
for the following reasons.
Consider the operator versions of the constraints $\mathcal{H},\underline{G},C$. 
These, when quantized, are smeared with test functions $N,A_0,\Lambda$ respectively and 
have the form $\hat{H}=\sum_v N(v)\hat{H}_v$ etc. In the case of $C_a$ one simply uses the 
operator in \eref{cond}. Now consider the operator matrix $\hat{M}_{ij}(v,v')$ - which could be 
interpreted as the operator version of $M^{(P)}_{ij}(x,y)$ - defined with the help of the commutators of 
the constraint operators: $\hat{M}_{12}=[\hat{H}_v,\hat{\underline{G}}_{v'}]$ etc. The question is whether the 
inverse of this matrix - defined via the condition $\sum_{v''}\hat{M}_{ik}(v,v'')(\hat{M})^{-1}_{kj}(v'',v')=\delta_{ij}\delta_{vv'}$ - 
actually exists. If it does, then with the help of this matrix we can define a Dirac commutator - analog 
of the Dirac-bracket - the following way:
\begin{eqnarray}
\hspace{-2cm}[\hat{O}_1,\hat{O}_2]_D=[\hat{O}_1,\hat{O}_2]-\nonumber\\
\hspace{-2cm}-\frac{1}{2}\sum_{v_1,v_2}([\hat{O}_1,\hat{C}_i(v_1)](\hat{M})^{-1}_{ij}(v_1,v_2)[\hat{C}_j(v_2),\hat{O}_2]-[\hat{O}_2,\hat{C}_i(v_1)](\hat{M})^{-1}_{ij}(v_1,v_2)[\hat{C}_j(v_2),\hat{O}_1]),\label{qdirac}
\end{eqnarray}
where the $\hat{C}_i$ are the constraint operators. 
What now one has to do is to replace the commutators with this Dirac commutator. Further more one has to 
modify the action of the momentum operators. To see why, consider first a dust network state 
\begin{eqnarray}
|D(U)>_{\gamma,\vec{\lambda}}:=\prod_{i=1}^{N}(U(\lambda_i,v_i))
\end{eqnarray}
which can be interpreted if the operator $\prod_{i=1}^{N}(U(\lambda_i,v_i))$ acted on the vacuum state $|0>$.
If we have first class constraint algebra we have the following identity:
\begin{eqnarray}
\hat{P}|D(U)>_{\gamma,\vec{\lambda}}=[\hat{P},\prod_{i=1}^{N}(U(\lambda_i,v_i))]|0>
\end{eqnarray}
After gauge fixing this identity is the key to define the new momentum operator $\hat{P}_D$:
\begin{eqnarray}
\hat{P}_D|D(U)>_{\gamma,\vec{\lambda}}:=[\hat{P},\prod_{i=1}^{N}(U(\lambda_i,v_i))]_D|0>
\end{eqnarray}
The new electric flux operator $\hat{E}(S)_D$ can be defined the same way. The only thing one has to do 
is replace the momentum operators in the constraints and use Dirac commutators instead of usual commutators (note 
that the gravitational momentum operator does not change since the Poisson-bracket of $E_a^i$ is zero with all four 
second class constraints).\\
The critical part of the above construction is the existence and uniqueness of $(\hat{M})^{-1}_{ij}(v,v')$. But 
even if it would exist, there is the other question whether \eref{qdirac} is really the operator version of the 
Dirac-bracket? The answers to these questions are the requirement that the above construction works.\\ 
If the above operator exists then the anomalies of the constraint algebra are removed. First let us focus on the 
gravitational variables. Since the gravitational gauge and the diffeomorphism constraints are first class, and the 
momentum operator for the canonical momenta $E_a^i$ does not change after gauge fixing, there will be no 
gravitational anomalies in the theory. Other anomalies will not appear since by construction $\{C_i,C_j\}_D=0$ for 
all constraints $C_i$ and if \eref{qdirac} is the operator version of the Dirac-bracket, we obtain 
\begin{eqnarray}
\hspace{-2cm}[\hat{C}_i,\hat{C}_j]_D=[\hat{C}_i,\hat{C}_j]-\nonumber\\
\hspace{-2cm}-\frac{1}{2}\sum_{v_1,v_2}([\hat{C}_i,\hat{C}_k(v_1)](\hat{M})^{-1}_{kl}(v_1,v_2)[\hat{C}_l,\hat{C}_j]-[\hat{C}_j,\hat{C}_k(v_1)](\hat{M})^{-1}_{kl}(v_1,v_2)[\hat{C}_l,\hat{C}_i])=\nonumber\\
\hspace{-2cm}=[\hat{C}_i,\hat{C}_j]-\frac{1}{2}([\hat{C}_i,\hat{C}_j]-[\hat{C}_j,\hat{C}_i])=0
\end{eqnarray}
so there are no anomalies (the above commutator does not impose a new constraint since it is identically zero). This 
is also true for first class constraints that have vanishing Poisson-bracket with all constraints, since then 
the corresponding operators will have zero commutator, thus the previous expression is also zero. The only problem is 
the case of first class constraints that have non-zero Poisson-bracket with the constraints since then the structure 
constants will appear in the Dirac-bracket. This is problematic in the quantum theory since the structure constants 
may become operators which could cause anomalies. In all cases factor ordering ambiguities occur becuse we have terms 
that contain the product of three non-trivial operators, but these do not cause inconsistencies but only change the 
results of the theory. 

\section{Mass}

Note that in this theory mass is a parameter, in fact a coupling constant which
couples the scalar field, the Yang-Mills field and gravity. In the classical
Hamiltonian analysis one can make the following rescaling: 
$\pi\to \pi/m,\ \phi\to m\phi,\ \underline{A}_a\to\underline{A}_a/m,\ \underline{E}_a\to m\underline{E}_a$, 
which is a canonical transformation. But in the quantum regime, this parameter enters
the Hamiltonian in a non-trivial way. In this sense it is very similar to the
Immirzi parameter of the pure gravitational case. In the latter case the Hawking-entropy
provided a tool that helped fix this parameter (\cite{revt},\cite{reval}), so there is a chance 
that with a similar method one might be able to make predictions on the value of m.\\
Another way would be to define propagators in loop quantum gravity, since the poles of 
the propagators could be interpreted as mass. But so far the question of time remains unsolved 
in the theory, leaving this idea for future research. None the less there are attempts which 
could provide a solution of the problem of time, see e.g. \cite{time} and references therein.
But without further input, mass is an undefined parameter of the theory 
which has to be given from experiments.

\section{Summary and outlook}

In this paper we investigated the Proca-field in the framework of loop quantum gravity. It 
turned out that the tools developed in this theory can be applied to the Proca-field if one 
introduces a scalar field and rewrites the Lagrangian as \eref{sympl}. But this scalar field 
is quite different than the one used in spontaneous symmetry breaking, since there is no need 
for scalar mass term and it restores symmetry rather breaking it. The role of this field is 
actually to make the constraint algebra first class. After rewriting the Lagrangian, the 3+1 
decomposition and the quantization was straightforward. The resulting Hamilton operator is 
well defined and diffeomorphism covariant. We provided a method to calculate the matrix elements of  
this Hamiltonian and some special solutions, which will may be a starting point on the interpretation 
of the theory. We showed that the parameter m is actually a coupling constant and is very 
similar to the Immirzi parameter. We also showed how to introduce a gauge fixing both at the classical 
and quantum level which gives 
back the original theory. This method could be useful for quantizing general systems with second 
class constraint algebra. \\
Three major questions remain. The first is the question of the mass. In our description we treated 
it as a given parameter and did not obtain any condition which could have given it fixed values. 
Research in this direction could give us values for m which could prove useful testing Loop 
Quantum Gravity.\\
The second question is closely related to the first and it is the relationship with theories 
with spontaneous symmetry breaking. This would give some more understanding of the Higgs mechanism 
on one hand and gauge fixing on the other.\\
The third is that one should give a more precise and more general method for gauge fixing in order to use it in other 
cases. For example the gauge fixing used in the case of the Proca-field was simple in the sense that 
it did not affect the gravitational part of the theory (it was diffeomorphism invariant etc.). But one 
could imagine gauges that break diffeomorphism invariance where the construction given here cannot be 
applied.

\section*{Acknowledgements}

This work was supported by the OTKA grant No. TS044839. The author would also like to 
thank the Polish and Hungarian Academy for financial support and 
the valuable consultations with Jerzy Lewandowski, Jerzy Kijowski, Istvan Racz and Gyula 
Bene.

\end{document}